\documentclass[11pt,a4paper]{article}
\usepackage{jheppub}
\usepackage{caption}
\usepackage{amsmath,amsfonts,amstext,amssymb,amsthm}
\usepackage{epigraph}
\usepackage{mathtools}
\usepackage{epstopdf}
\usepackage{subcaption}
\usepackage{float}
\usepackage{parskip}
\usepackage{verbatim}
\usepackage{comment}
\usepackage{enumitem}
\usepackage{hyperref}
\usepackage{dsfont}
\usepackage{shuffle}
\usepackage{graphics}
\usepackage{graphicx}% Include figure files
\usepackage{dcolumn}% Align table columns on decimal point
\usepackage{bm}% bold math
\usepackage{epsfig}
\usepackage{multirow}
\usepackage{dcolumn}% Align table columns on decimal point
\usepackage{graphicx,epsfig}%
\usepackage[mathscr]{euscript}
\usepackage{notoccite}
\usepackage{xhfill}
\usepackage{makecell}
\usepackage[perpage]{footmisc}
\usepackage{diagbox}

\begin{document} 

\author[a,1]{A. B. Kormilitzin,\note{Corresponding author.}}
\author[b]{K. E. A. Saunders,}
\author[b]{P. J. Harrison,}
\author[b]{J. R. Geddes,}
\author[a]{T. J. Lyons}

% The "\note" macro will give a warning: "Ignoring empty anchor..."
% you can safely ignore it.

\affiliation[a]{Mathematical Institute, University of Oxford, Andrew Wiles Building,\\ Woodstock Rd, Oxford OX2 6GG, UK}
\affiliation[b]{Department of Psychiatry, Oxford University, and Oxford Health NHS Foundation Trust,\\ Warneford Hospital, Oxford OX3 7JX, UK}

% e-mail addresses: one for each author, in the same order as the authors
\emailAdd{andrey.kormilitzin@maths.ox.ac.uk}
%\emailAdd{second@asas.edu}
%\emailAdd{third@one.univ}
%\emailAdd{fourth@one.univ}

\abstract{The classification procedure of streaming data usually requires various ad hoc methods or particular heuristic models. We explore a novel non-parametric and systematic approach to analysis of heterogeneous sequential data. We demonstrate an application of this method to classification of the delays in responding to the prompts, from subjects with bipolar disorder collected during a clinical trial, using both synthetic and real examples. We show how this method can provide a natural and systematic way to extract characteristic features from sequential data.}


\title{Application of the Signature Method to Pattern Recognition in the CEQUEL Clinical Trial}

\maketitle
\flushbottom

\section{Introduction}

The analysis of streaming data is one of the biggest challenges posed by the expansion of
digital healthcare and bioinformatics. A data stream is a sequence of data that arrives over time. Familiar examples are stock prices, sensor data from mobile devices, personal data from monitoring platforms and many more. 

The field of machine learning and data mining offers various frameworks for discovering
patterns, hidden information, and learning the functional dependencies in complex data. Most methods in machine learning require a good choice of characteristic features to learn functions or compute the posterior probabilities. Feature extraction and selection methods are numerous. Unfortunately, there is no clear consensus as to how analyse and extract features from heterogeneous sequential data. In this work we are aiming to introduce a novel framework for the analysis of data streams. Our approach is based on the mathematical theory called {\it rough paths theory}. This allows one to solve controlled differential equations driven by rough signals \cite{lyons1998differential} as well as to solve stochastic partial differential equations \cite{kloeden2012numerical,kusuoka2004approximation,lyons2004cubature}. The foundations of the proposed approach date back to the seminal work \cite{chen1957integration} on iterated integrals of multidimensional piecewise smooth paths. The core ingredient of the theory is {\it a signature}. The signature is a transformation of a path into the sequence of its coordinate iterated integrals, whereas the integrals obey particular algebraic properties.

Here we describe an application of the signature method to the analysis of longitudinal mood data from the clinical trial of treatments for bipolar depression. Bipolar disorder is a mental illness characterised by episodes of elevated mood (manic episodes) and periods of depression (low mood states) \cite{anderson2012bipolar}. Depressive episodes are longer in duration that periods of mania and are associated with long-term disability and a high risk of suicide \cite{judd2003prospective, judd2002long, goodwin2003suicide}. Effective treatments for bipolar depression are limited. One of the widely used medications is quetiapine, an atypical antipsychotic agent. Another recommended treatment is lamotrigine, an anticonvulsant.

CEQUEL (Comparative Evaluation of QUEtiapine-Lamotrigine combination versus quetiapine monotherapy in people with bipolar depression) was a double blind randomised placebo controlled parallel group trial  of lamotrigine plus quetiapine versus quetiapine monotherapy in patients diagnosed with bipolar disorder currently suffering from a depressive episode.   \cite{geddes2016comparative}. %It also explored the effects of folic acid supplementation in a 2 factorial design.

\begin{figure}[htbp]
\centering
  \begin{subfigure}[b]{0.45\textwidth}
  \centering
    \includegraphics[width=\textwidth]{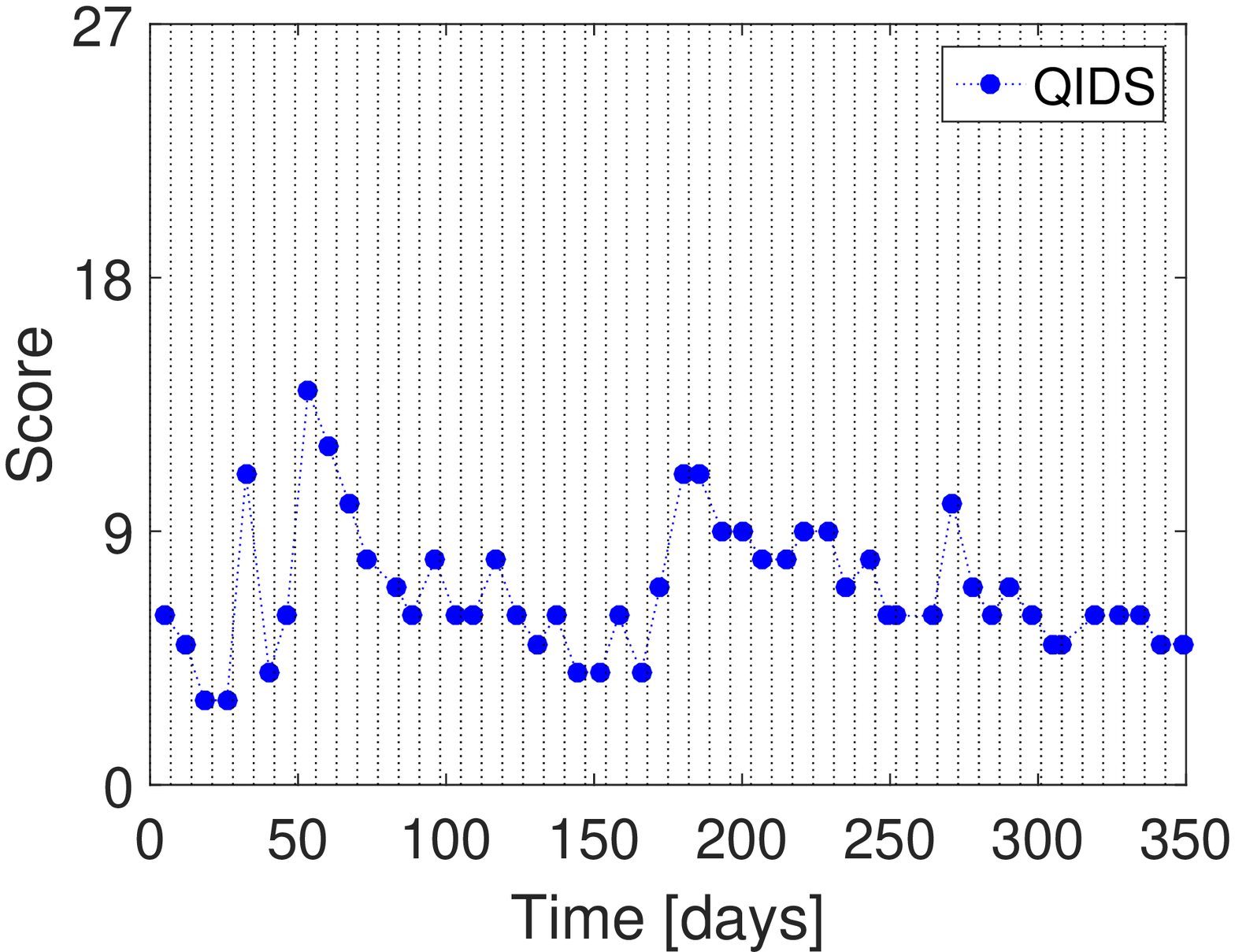}
    \caption{QIDS-SR$_{16}$}
    \label{fig:typicalQIDS}
  \end{subfigure}
 \hfill
  \begin{subfigure}[b]{0.5\textwidth}
  \centering
    \includegraphics[width=\textwidth]{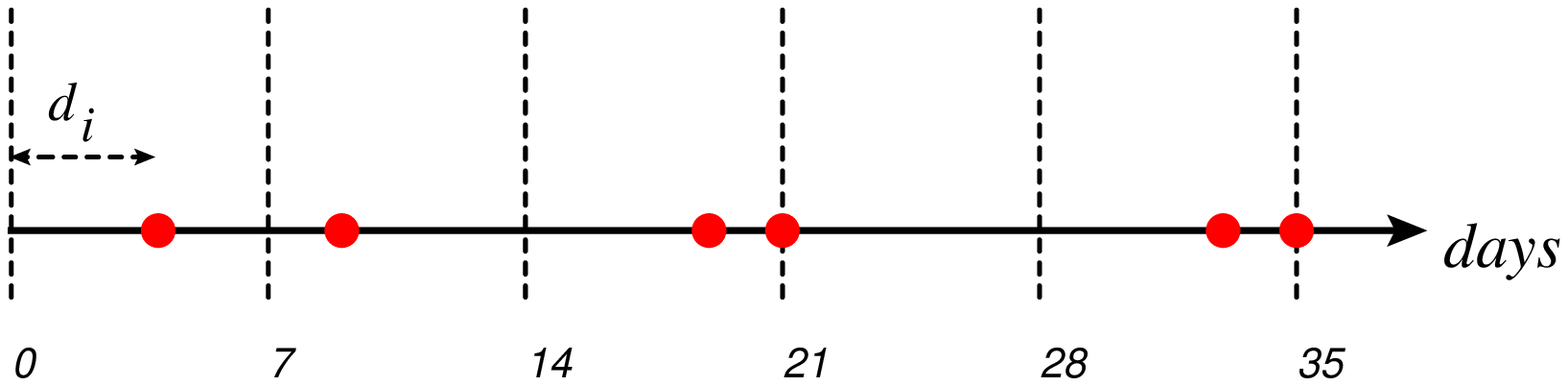}
    \caption{Delays}
    \label{fig:delaysOld}
  \end{subfigure}
  \caption{Left panel: examples of two rating scales QIDS-SR$_{16}$ (blue line) and the AMSR (red line). Right panel: Response delays $d_i$ of subjects. The dashed vertical lines correspond to the time stamp of the prompt text and the red dots correspond to time stamps of actual responses. The sequence of delays in this example is: $d = (4,2,5,0,5,0)$.}
\end{figure}

The primary outcome measure for the trial was improvement at 12 weeks in patient-reported depressive symptoms using the the 16 item Quick Inventory of Depressive Symptomatology - Self Report (QIDS-SR$_{16}$) \cite{rush200316}. %Symptoms of mania were assessed using the 5-item Altman Self-Rating Mania Scale (ASRM) for manic symptoms \cite{altman1997altman}.
The resulting overall scores are represented by integer numbers and lies in the range between 0 and 27 for QIDS-SR$_{16}$. %and between 0 and 25 for ASRM. 
Subjects were encouraged to submit their QIDS-SR$_{16}$ % and ASRM scores 
on a weekly basis, and were prompted by text or e-mail sent using the True Colours platform (TC)\footnote{https://truecolours.nhs.uk/demo/documents/PatientGuide.pdf}. Missing data correspond to absence of response within one week interval between the reminders. Typical self-reported data from subjects is presented in Fig.~\ref{fig:typicalQIDS}.

\noindent

The original data analysis \cite{geddes2016comparative} reported a significant improvement in depressive symptoms in patients randomised to lamotrigine compared with placebo.In our reanalysis of the data we focus on timestamps of the responses rather than the actual rating scale scores and examine the regularity of responses in the different treatment groups. The objective data were delays, defined as time intervals between the prompts and the actual response. Delays are schematically depicted in Fig.~\ref{fig:delaysOld}. These represent the time it took for participants to complete their mood ratings following the receipt of a prompt to do so. Psychomotor retardation is a common symptom of depression \cite{bennabi2013psychomotor,buyukdura2011psychomotor,american2013diagnostic} and these delays are likely to be representative of this impairment. Shortening of delays would predicted to occur as depressive symptoms improved. Given that mood lability, in addition to depression and mania, characterizes bipolar disorder \cite{harrison2016innovative}, we were also interested to determine the regularity in response timings.

We formulated the problem as a binary classification problem: is there a difference between two treatment groups $``0"$ (placebo) and $``1"$ (lamotrigine) conditioned on the distribution of delays.

\section{Methods}

One may think of the signature as a transformation of data into a sequence of real numbers (signature terms), which summarise the information hidden in data. These signature terms are then used as a systematic set of features of the data for any machine learning task. Intuitively, this method is similar to spectral methods (for example, the Fourier transform), where the iterated integrals of data serve as basis functions, but quite different conceptually. Recently, the signature framework for the time-series models \cite{levin2013learning} has been proposed and successfully applied to financial data streams \cite{gyurko2013extracting}. In the following sections we will elaborate on this approach and provide the detailed mathematical derivations.

\subsection{Continuous Paths and their relation to raw sequential data}

\begin{figure}[htbp]
\vskip 0.0in
\begin{center}
\centerline{\includegraphics[width=\columnwidth]{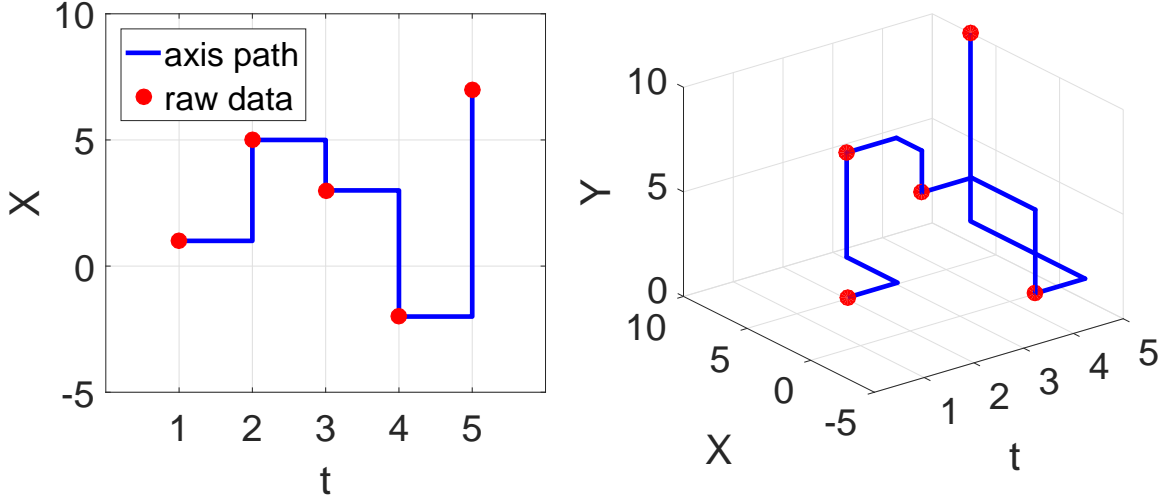}}
\caption{Examples of embedding of the raw discrete data into continuous paths: in two dimensions (left panel) and in three dimensions (right panel). The data set is given by: $t=\{1,2,3,4,5\}$, $X=\{1,5,3,-2,7\}$, $Y=\{2,7,5,1,10\}$.}
\label{fig:pathsFromData}
\end{center}
\vskip -0.2in
\end{figure}

Consider a collection of $N$ $d$-tuples $X = \{X_{1_i},X_{2_i},...,X_{d_i}\}_{i=1}^N$, where each $X_{k_i}$ represents an individual one-dimensional sequential data. We embed the discrete set of data into a continuous path by piecewise interpolation methods. For example, consider a set of pairs $\{(t_i,X_i)\}$ and a set of triples $\{(t_i,X_i,Y_i)\}$.  In Fig.~\ref{fig:pathsFromData}, the left panel shows the stairstep piecewise ({\it axis path} \cite{lyons2011inversion}) embedding of the pairs $X=\{t_i,X_i\}_{i=1}^{N=5}$ into a path, and the right panel shows a path constructed from the set of triples $X=\{t_i,X_i,Y_i\}_{i=1}^5$ in three-dimensional space.

\noindent
Another important type of embedding of discrete data streams is the {\it Lead-Lag} transformation. It accounts for the relationship between the subsequent elements of the streams, which results in revealing the cross correlated information hidden in data. Formally, the lead-lag transformation of the stream $\{X\}_{i=0}^N$ is defined separately for the lead and the lag streams:
\begin{eqnarray}
    X_{2n+2}^{lead} &=& X_{2n+1}^{lead} = X_{n+1};\;\;\forall n \in \{0,...,N-1\}\\\nonumber
    X_{2n}^{lag} &=& X_{2n+1}^{lag} = X_{n+1}\nonumber
\end{eqnarray}
The lead-lag transform of the stream $\{X\}_{i=0}^N$ is defined by the two concatenated streams:
\begin{equation}\label{eq:leadLag}
    \{X^{lead-lag}\}_{i=0}^{2N} = \{(X^{lead}_i, X^{lag}_i)\}_{i=0}^{2N}
\end{equation}
The transformation of data streams into continuous paths in higher-dimensional space is the first step in our approach, followed by the computation of coordinate iterated integrals of continuous paths.

%%%%%%%%%%%%%%%%%%%%%%%%%%%%%%%%%%%%%%%%%%%%%%%%%%%%%%%%%%%%%%%%%%%%%%%%%%%

\subsection{The Signature of a Path}

The signature of a path is a sequence of its coordinate iterated integrals \cite{lyons2007differential}. Let the $X$ be a  path in $\mathbb{R}^d$ parametrized on interval $[0, T]$ with finite $p$-variation for some $p < 2$. The iterated integrals of order $k$ with multi-index $I=(i_1,i_2,...,i_k)$ of the path $X$ are defined as:
\begin{equation}\label{eq:sigTerms}
    X^I = \int_{0<u_1<u_2<...<u_k<T}dX^{i_1}_{u_1}dX^{i_2}_{u_2}...dX^{i_k}_{u_k}.
\end{equation}
The signature of $X$ denoted by $S(X)$ is the infinite sequence of all coordinate iterated integrals
\begin{equation}\label{eq:infSig}
    S(X) = \left(1,X^{(1)},X^{(2)},...,X^{(d)},X^{(1,1)},X^{(1,2)},...\right),
\end{equation}
where the multi-index $I$ consists of all combinatorial combinations of $i_j\in\{1,2,...,d\}$ for $j=1,2,..k$. For practical application of the signature transformation, we normally truncate the infinite sequence (\ref{eq:infSig}) at some level $L$ by taking only first terms of the full signature. In most applications it is enough to truncate the sequence at level $L=4$, or even $L=2$. The signature transformation and its iterated integrals obey several important mathematical properties:
\begin{itemize}
    \item Invariance of signature to re-parametrisation of paths $X$.
    \item Any product of iterated integrals could be expressed as a linear combination of the higher order iterated integrals.
    \item The signature of the concatenated path is given by a tensor product of signatures of each individual paths.
    \item Signature is invariant to presence of tree-like paths.
\end{itemize}

\begin{figure}[htbp]
\vskip 0.2in
\begin{center}
\centerline{\includegraphics[width=0.7\columnwidth]{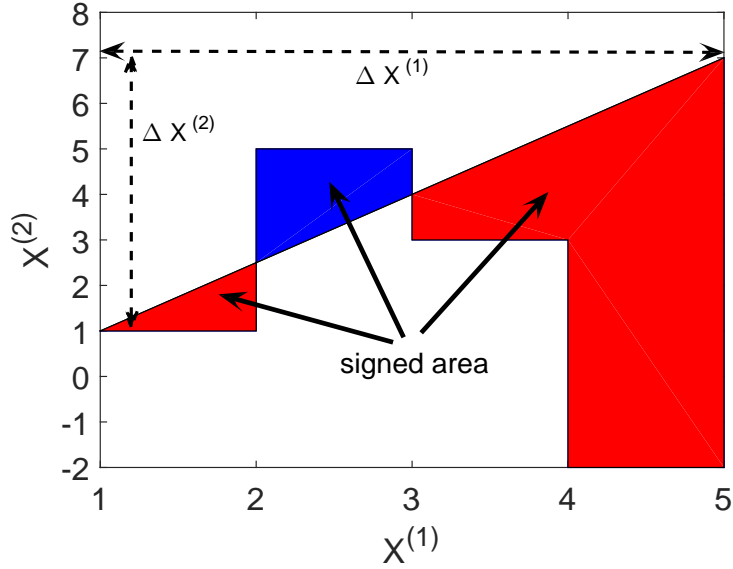}}
\caption{Geometrical interpretation of the low order signature terms. Two total increments $\Delta X^{(1)}$ and $\Delta X^{(2)}$ and the signed area are shown.}
\label{fig:sigTermsGeom}
\end{center}
\vskip -0.2in
\end{figure}

There is a simple, intuitive interpretation of the low order signature terms. Consider a path which takes values in $\mathbb{R}^2$, for example the one depicted on the left panel in Fig.~\ref{fig:pathsFromData} parametrised as $\{(X^1_i,X^2_i\} = \{(t_i,X_i)\}$.  Its single iterated integral has a meaning of the {\it increment} of a path. Following the definition (\ref{eq:sigTerms}):
\begin{equation}
    X^{(k)} = \int_{0<u_1<T}dX^k_{u_1} = X^k_T - X^k_0 = \Delta X^k;\;\;k = 1,2 
\end{equation}
The double-iterated integrals of paths have meaning of {\it signed area} enclosed by the path in $\mathbb{R}^2$ and a straight chord connecting the endpoints of the path. The relation between the double-iterated integrals and the area is easily deduced from the Green's theorem, namely:
\begin{eqnarray}\label{eq:sigLowTerms}
    X^{(1,2)} &=& \int_{0<u_1<u_2<T}dX^1_{u_1}dX^2_{u_2}\nonumber\\
    X^{(2,1)} &=& \int_{0<u_1<u_2<T}dX^2_{u_1}dX^1_{u_2}\nonumber\\
    A &=& \frac{1}{2}\left(X^{(1,2)} - X^{(2,1)} \right)
\end{eqnarray}
The geometrical meaning of these terms are presented in Fig.~\ref{fig:sigTermsGeom}. The positive and the negative areas are coloured in red and in blue respectively. The total area is given by the sum of its positive and negative parts and is exactly equal to $A$ in (\ref{eq:sigLowTerms}).

\noindent

It becomes less obvious to interpret the higher order iterated integrals, although one may deduce intuitively that $k$-fold iterated line integral corresponds to some hypervolume in $\mathbb{R}^k$.

The important property of the lead-lag transform defined in (\ref{eq:leadLag}) is that the signed area between the lead-stream and the lag-stream corresponds to the quadratic variation of the data stream $X$.

The algebraic properties of signature terms are made transparent in the space of formal series of tensors $T((E))$ of Banach space $E$. The space of sequences defined as
\begin{equation*}
    T((E)) = \left\{ a=(a_0,a_1,...)| \forall n\ge0, a_n \in E^{\otimes n} \right\},
\end{equation*} with \begin{equation*} 
{\bf a}+{\bf b} = (a_0+b_0,a_1+b_1,...)\end{equation*} and \begin{equation*}{\bf a}\otimes {\bf b} = \left(a_0 \otimes b_0, a_0\otimes b_1 + a_1\otimes b_0, ..., \sum_{k=0}^n)a_k\otimes b_{n-1},...\right). \end{equation*} For example, a product of two first order signature terms is given by the linear combination of second order terms: $X^{(1)} X^{(2)} = X^{(1,2)} + X^{(2,1)}$. The linearisation of signature terms due to shuffle product of iterated integrals has a potential for practical application of linear regression to non-linear problems.

%%%%%%%%%%%%%%%%%%%%%%%%%%%%%%%%%%%%%%%%%%%%%%%%%%%%%%%%%%%%%%%%%%%%%%%%%%%

\subsection{The Signature Based Methodology in Machine Learning}

It has been shown in \cite{hambly2010uniqueness} that any data stream of finite length is represented by its signature and Poincare-Birkhoff-Witt theorem ensures that the coordinate iterated integrals are a good choice for parametrisation of the polynomials on this space. This theorem underpins our motivation to use the signature terms as a faithful feature set of data streams. The application of the signature transformation of data streams to machine learning problems is straightforward and has already demonstrated its success \cite{levin2013learning,gyurko2013extracting,yang2015character,chevyrev2016primer,kiraly2016kernels}. One of the best results to date for use of this method has been in the recognition of Chinese characters \cite{yin2013icdar}. %, where the first place in the international competition went to an indicvidual who use the signature method \cite{graham2013sparse} for feature extraction together with state-of-the-art deep learning methods.

Here we outline the general approach of the signature method to various machine learning problems. The workflow for feature extraction using the signature method is depicted in the Fig.~\ref{fig:sigWorkFlow} and contains several important steps:
%THIS FIGURE HAS SPELLING ERROR IN IT! FEATURE IS WRONG

\begin{figure}[htbp]
\vskip 0.2in
\begin{center}
\centerline{\includegraphics[width=\columnwidth]{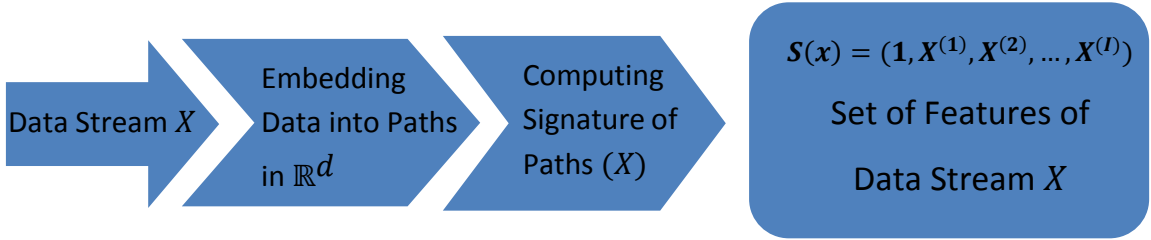}}
\caption{The workflow of the feature extraction using the signature method.}
\label{fig:sigWorkFlow}
\end{center}
\vskip -0.2in
\end{figure} 
\noindent
\begin{itemize}
    \item Embed discrete data into continuous paths. 
    \item Use the lead-lag transform to account for the quadratic variability in data.
    \item Compute the coordinate iterated integrals up to specified level of truncation $L$.
    \item Standardise the signature terms by means of the first and the second statistical moments.
    \item Use the resulting feature set for machine learning tasks.
\end{itemize}
The number of the signature terms grows very fast as function of the dimensionality of the path and the truncation level, thus for the supervised learning applications, both regression and classification, one should use regularisation techniques (LASSO or Ridge) in order to shrink the initial feature set. One may prefer to use the hybrid combination of both techniques - the elastic net regularisation \cite{tibshirani1996regression,zou2005regularization} to account for the multi-colinearity in the feature matrix.

\subsection{Dealing with Missing Data}

\begin{figure}[htbp]
\vskip 0.2in
\begin{center}
\centerline{\includegraphics[width=0.8\columnwidth]{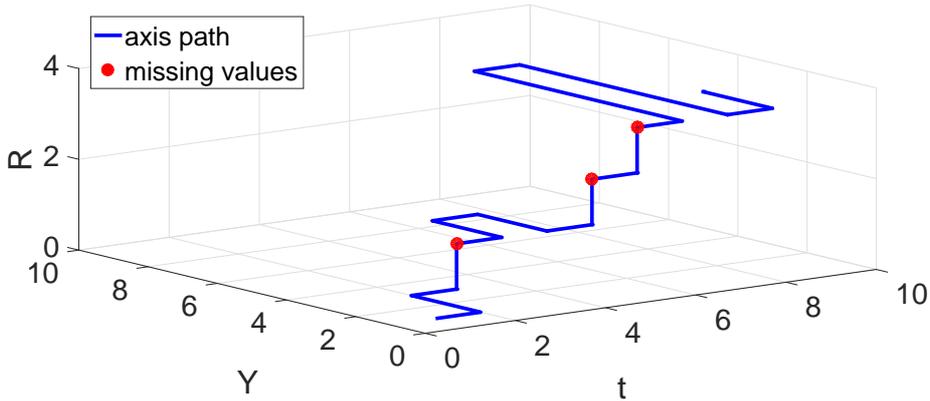}}
\caption{Example of embedding of data with missing values into a single path. The red dots represent unobserved data.}
\label{fig:missingData}
\end{center}
\vskip -0.2in
\end{figure}

One  property of the path representation of data is its ability to incorporate the missing data. The absence of data in $\mathbb{R}^d$ is treated as a presence of data in higher dimension $\mathbb{R}^{d+1}$. Introducing a Boolean {\it indicator variable} which has the same dimension as the data stream with missing data and marks by ``1" if the expected value is missing. For example, consider a one-dimensional data stream of values: 
\begin{eqnarray}\label{eq:missing_data_0}
Y_j\;&=&\;\{1,3,\star,5,3,\star,\star,9,3,5\}\\
R_j\;&=&\;\{0,0,1,0,0,1,1,0,0,0\}\nonumber
\end{eqnarray}
where the symbol ``$\star$" denotes missing value at expected time point. The idea is to create a two dimensional path from this data. The evolution of the path is from the starting point towards the end and might be seen as propagation in three dimensional space. First two dimensions are ``observed data" and ``missing data", while the third direction corresponds to time. Every time we have a missing point, we jump from ``observed" to ``missing" dimension and filling in the missing place with the same value as seen before (a.k.a {\it feed forward} method). The intuition is simple: unless we have new information about the data, we continue walking along the ``missing" axis direction, which is different than the ``observed" data direction. The resulting path is given by (\ref{eq:missing_data}):
\begin{align}\label{eq:missing_data}
	\tilde{Y}_{j} = \{(0,1,0),&(1,3,0),(2,{\bf 3},1),(3,5,0),(4,3,0)\\
	&(5,{\bf 3},1),(6,{\bf 3},1),(7,9,0),(8,3,0),(9,5,0)\}.\nonumber
\end{align}
Here we introduce an auxiliary integer-time parametrization vector $t\;=\;(0,1,2,\dots,9)$ to account for the time propagation and create a path in three-dimensional space as depicted in Fig.~\ref{fig:missingData}. Red points represent the missing values with coordinates $(t_j,Y_j,R_j) = (t_j,0,1)$. The path propagates from $t=(0,1,\dots,9)$ and the observed points lie in plane $(t_j\,Y_j)$, while in the presence of missing value, the path goes to $(t_j\,R_j)$.

\noindent

This approach allows to treat data streams with unobserved data at the same footing as complete data streams, which demonstrates another conceptual advantage of the signature framework. The construction of lifted paths by accounting for various variables lies in the core of the signature approach and allows one to build a systematic framework for analysis of heterogeneous sequential data.

\subsection{The CEQUEL Data}

The CEQUEL data set contains 202 subjects randomly assigned to different treatment groups. %We sought to replicate the primary research question to see whether any between-group difference could be identified within the first 12 weeks of the study. 
We focused on the responses made during the first 12 weeks of the trial, to reflect the fact that the primary outcome of CEQUEL was defined as QIDS score at the 12 week time point. Participants were subselected based on their response rate, where no more than two consecutive missing weeks are allowed, and available information about the timing of messages sent to prompt a response. Unfortunately, due to technical difficulties with the TC platform during the CEQUEL trial, the timestamps of the prompt messages were not recorded for a significant number of participants. The final subset of participants that matched our selection criteria were significantly reduced from 202 to 29; 11 were in the lamotrigine group and 18 in the placebo group.

\section{Results}

We used the signature method to extract features from the time series of delays $\{d_i\}$ and then we learnt a binary classifier to discriminate between the two groups $``0"$ (placebo) and $``1"$ (lamotrigine). 

\begin{figure}[htbp]
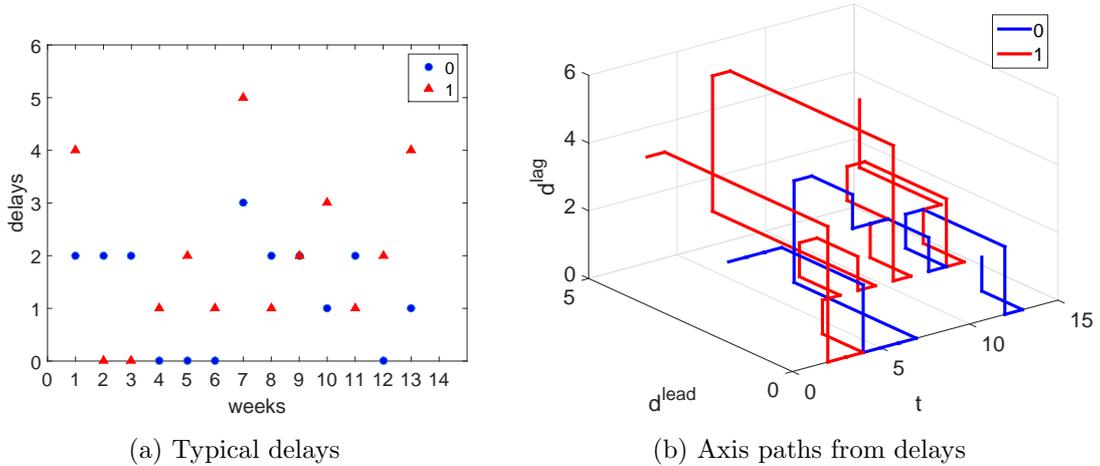

\centering
\begin{subfigure}[b]{0.45\textwidth}
  \centering
  \includegraphics[width=0.9\linewidth]{Fig_6a}
  \caption{Typical delays}
  \label{fig:delaysTSexample}
\end{subfigure}%
\begin{subfigure}[b]{0.55\textwidth}
  \centering
  \includegraphics[width=0.9\linewidth]{Fig_6b}
  \caption{Axis paths from delays}
  \label{fig:examplePath01}
\end{subfigure}
\caption{Left panel: The example of two randomly drawn subjects from the placebo ``0" and the lamotrigine ``1" group for 13 weeks of follow-up observations. The values $d$ of delays are defined in Fig.~\ref{fig:delaysOld}. Right panel: the example of embedding of points into axis paths from the discrete data $\{d_i\}$ shown in Fig.~\ref{fig:delaysTSexample}.}
\label{fig:test11}
\end{figure}

First we converted the discrete delay data $\{d_i\}$ into continuous axis paths denoted by $X$, together with the lead-lag transformation and using the integer-time component $\{t_i\} = (1,2,...,N)$:
\begin{equation}
    X = \left\{\left(t^{lead}_i,d^{lead}_i,d^{lag}_i\right)\right\}_{i=0}^N
\end{equation}
The example of typical distribution of the delays $\{d_i\}$ of subjects from the two groups are shown in Fig.~\ref{fig:delaysTSexample} and the embedding of these discrete data points into the axis path is demonstrated in Fig.~\ref{fig:examplePath01}.

\noindent
Then we extracted characteristic features by computing the iterated integrals of the path $X$ using the definition (\ref{eq:sigTerms}):
\begin{equation}
    S(X) = (1,X^{(1)},...,X^{I},...)
\end{equation}
where the $I$ is the multi-index $I=(i_1,i_2,...,i_k)$. The resulting features were standardised to have zero mean and unit variance. 

\subsection{Technical Details of the Classification Procedure}

The final data set of subjects prepared for the classification learning presented two main problems. It was both relatively small and imbalanced: $N_0 = 18$ versus $N_1 = 11$. The nature of imbalance was not intrinsic to underlying phenomena in data, but rather was an artifact of the data collection platform. Thus, we used a state-of-the-art approach to balance the number of subjects in both groups for more robust classification learning. The method we used was based on SMOTE (Synthetic Minority Over-sampling TEchnique)~\cite{chawla2002smote} with ADASYN implementation \cite{he2008adasyn} which outperforms a standard approach to over-sampling of the minority class with replacement. The core idea of the SMOTE method is to sample new synthetic features from the distribution of the original features. 

First we computed the features of the delay series using the signature method. We experimented with various truncation levels and due to the small sample size, we limited ourselves to level $L=4$. While higher truncation levels of the signature allow one to extract more information from the data (``a better resolution") improving the classification precision, in case of small data sets, increasing the level of truncation will simply overfit the data. Then we applied the SMOTE algorithm to balance the classes and we ended up with $N_0 = N_1$ = 18 subjects in each class. Effectively, we simulated features for the additional seven synthetic subjects from the lamotrigine group labeled by ``1".

We compared a performance of various classification algorithms on this data set to assess a robustness of our approach. Due to the linearisation property of the signature features, we used two linear classification models: the logistic regression and linear support vector machines (SVM) along with non-parametric the k-nearest neighbour (kNN) classifier. Since, the dimension of the feature space grows with the signature truncation depth $L$, we applied the elastic net regularisation algorithm to select the most relevant set of features. For each truncation level $L$ the resulting sets of features are:

\begin{eqnarray}\label{eq:sigLevels}
L = 2\;&:&\;\{X^{(2,1)},X^{(2,2)},X^{(2,3)}\}\\
L = 3\;&:&\;\{X^{(2,2)},X^{(2,1,2)},X^{(2,3,2)}\}\nonumber\\
L = 4\;&:&\;\{X^{(2,1,3)},X^{(2,1,1,2)},X^{(2,3,1,2)},X^{(3,3,3,2)}\}\nonumber
\end{eqnarray}
The number of relevant features which were used for the classification at each truncation level is highlighted in the Table~\ref{table:classificationResults}, where the total number of features at the same level is given in braces.

\begin{figure}[htbp]
\centering
\begin{subfigure}[b]{.5\textwidth}
  \centering
  \includegraphics[width=\linewidth]{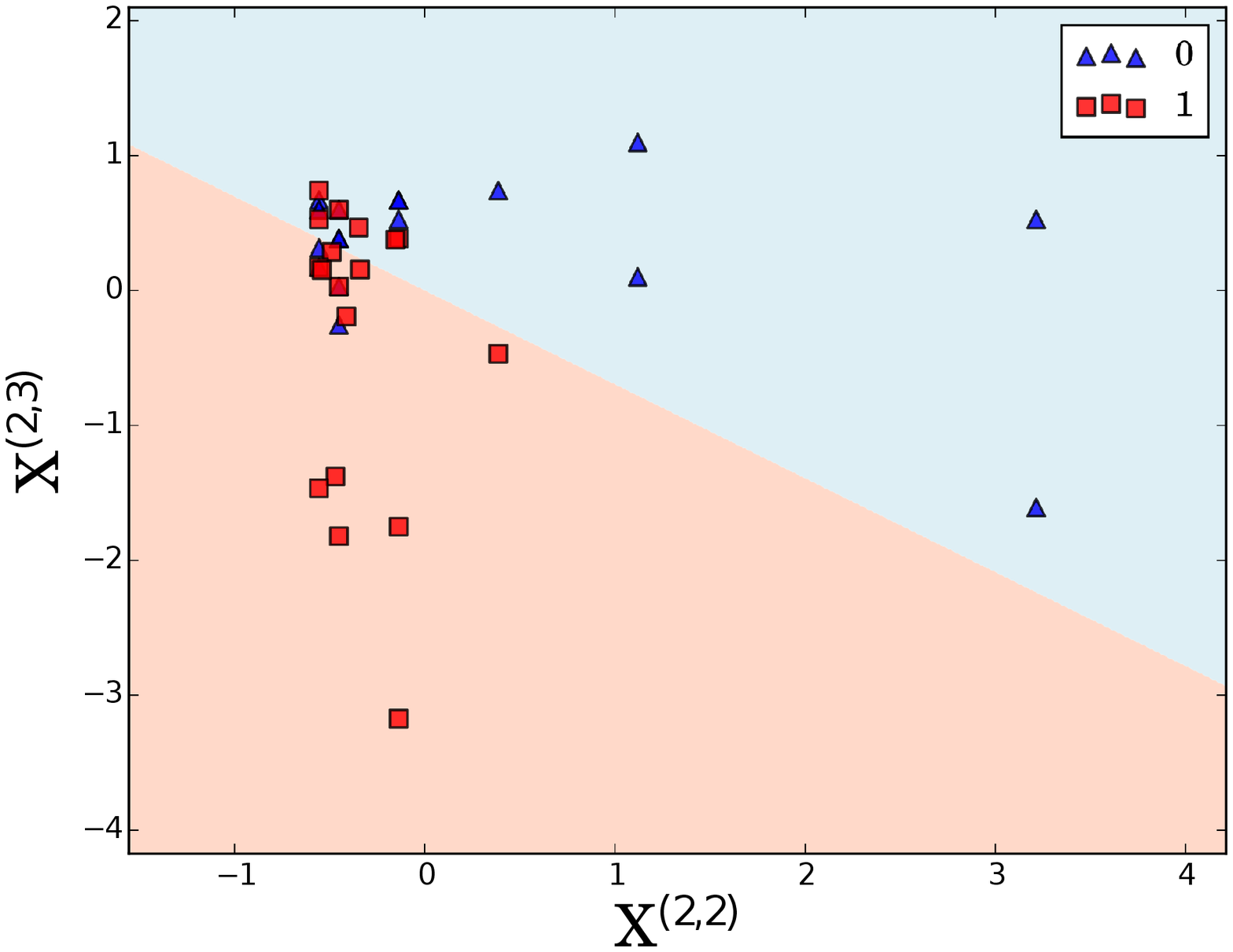}
  \caption{}
  \label{fig:feat_1}
\end{subfigure}%
%\begin{subfigure}[b]{.33\textwidth}
%  \centering
%  \includegraphics[width=\linewidth]{4_13}
  %\caption{Delays}
%  \label{fig:feat_2}
%\end{subfigure}
\begin{subfigure}[b]{.5\textwidth}
  \centering
  \includegraphics[width=\linewidth]{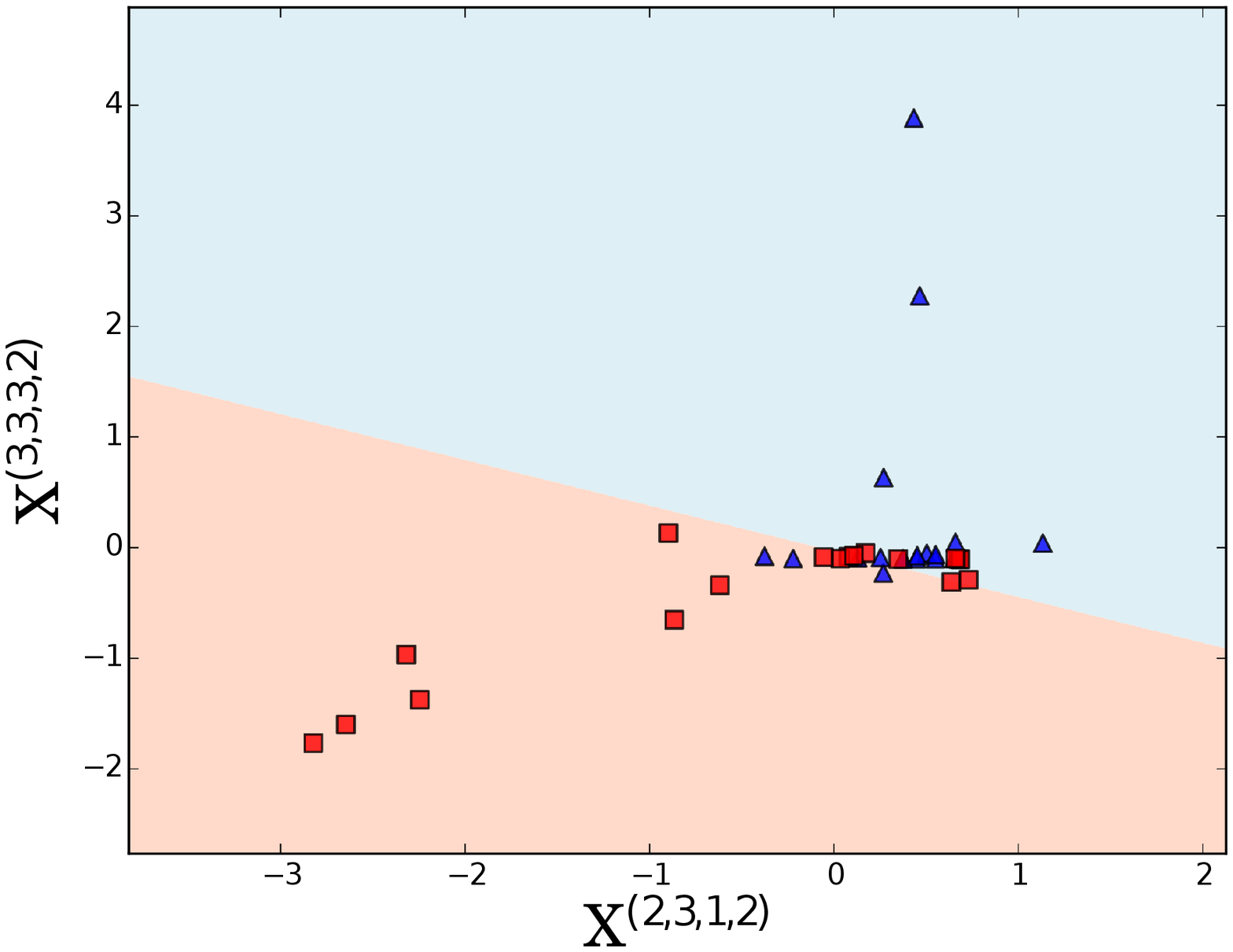}
  \caption{}
  \label{fig:feat_3}
\end{subfigure}%
\caption{Examples of projections of the classification decision hyperplane on selected features.}
\label{fig:featuresDelaysEN}
\end{figure}

Due to the small sample size, we used the nested $6$-fold cross-validation scheme with the stratified resampling method~\cite{kohavi1995study,cawley2010over} to tune hyperparameters of the classifiers and to assess the classification error. To estimate a performance of the classification we used several statistical metrics summarised in Table~\ref{table:classificationResults}. The projections of the selected features on two dimensional planes and the decision boundary between two groups are shown in Fig.~\ref{fig:featuresDelaysEN}.

\noindent
%%%%%%%%%%%%%%%%%%%%%%%%%%%%%%%%%%%%%%%%%%%%%%%%%%%%%%%%%%%%%%%%%%%%%%%%%%%%%%%%
%%%%%%%%%%%%%%%%%%%%%%%%%%%%%%%%%%%%%%%%

\begin{table}[htbp]
\centering
\begin{tabular}{|l|c|c|c|c|c|c|c|c|c|}\hline
Classifier &\multicolumn{3}{|c|}{Logistic regression} &\multicolumn{3}{|c|}{SVM} &\multicolumn{3}{|c|}{kNN}\\\hline\hline
Signature depth & {\bf L=2} & L=3 & L=4 & {\bf L=2} & L=3 & L=4 & {\bf L=2} & L=3 & L=4\\\hline\hline
number of features & {\bf 3} (7) & {\bf 3} (23) & {\bf 4} (74)& {\bf 3} (7) & {\bf 3} (23) & {\bf 4} (74)& {\bf 3} (7) & {\bf 3} (23) & {\bf 4} (74)\\\hline
sensitivity & 0.71 & 0.67 & 0.67 & 0.67 & 0.55 & 0.72 & 0.72 & 0.55 & 0.67 \\
specificity & 0.78 & 0.84 & 0.83 & 0.89 & 0.89 & 0.89 & 0.67 & 0.94 & 0.89 \\
accuracy    & 0.72 & 0.75 & 0.72 & 0.81 & 0.67 & 0.75 & 0.75 & 0.75 & 0.72 \\
f1-score    & 0.70 & 0.72 & 0.67 & 0.77 & 0.58 & 0.73 & 0.76 & 0.71 & 0.67 \\
AUC         & 0.78 & 0.83 & 0.83 & 0.85 & 0.80 & 0.78 & 0.73 & 0.77 & 0.81 \\
Cohen's kappa       & 0.50 & 0.50 & 0.50 & 0.56 & 0.44 & 0.61 & 0.39 & 0.28 & 0.56\\\hline
\end{tabular}
\caption{Statistical summary of the classification quality of various learning algorithms.}
\label{table:classificationResults}
\end{table}
\noindent

We compared a dependence of the classification accuracy on various signature truncation levels $L$ (\ref{eq:sigLevels}). It is clear from the Table~\ref{table:classificationResults} that there is no substantial gain in the classification performance with increasing $L$. In such cases the simplest model with minimal truncation level is preferred. The meaning of this result is that the signature transformation at minimal level is able to capture the essential structure in raw data and the higher order iterative integrals do not reveal any additional information.

\subsection{Computational Considerations}

The computation of the signature terms from discrete data points has been performed with the help of the {\it CoRoPa} project \cite{gyurko2008rough,WinNT} where the algorithms are implemented in C++ programming language and with a special python wrapper for computation of the inner product of the signature terms.

\section{Discussion and Conclusions}
\label{discussConclus}

We have described a novel systematic framework for heterogeneous sequential data analysis and feature extraction method, applied to analysis of data collected remotely during a clinical trial of treatment for depression in bipolar disorder. The characteristic set of features arise naturally in our approach and have the interpretable geometrical meaning as functions of data. The algebraic properties of iterated integrals linearise the non-linear polynomials of integrals, allowing us to use the linear regression models to learn functional dependencies in data. The feature set constructed from the signature terms could be used for any types of machine learning problems. The conceptual flexibility of the representation of data as paths $\mathbb{R}^d$ space broaden the application area of the signature method far beyond the typical sequential data. The signature approach does not involve any intrinsic parameters, serving as a good candidate for Bayesian learning.

From the uniqueness property of the signature it follows, that the terms $S^I$ are faithful representation of the original data, and locally approximate arbitrary well the functions of data. These properties make the signature method a legitimate approach to representation learning.

The numerical results of the classification learning demonstrated that the two treatment groups could be distinguished with approximately 75 percent accuracy (cf. Table~\ref{table:classificationResults}) on the basis of the signature transformation of the delays between the prompts and the responses. To combat the imbalance of the original data set, we applied the SMOTE algorithm to generate synthetic features for the less prevalent group (lamotrigine) with successive resampling of the synthetic data to ensure the robustness of the result.

%lead-lag relationship between prompts and responses.

%The application of the signature method to classification of participants in the CEQUEL trial revealed that those two treatment groups could be distinguished with 80 percent accuracy on the basis of the lead-lag relationship between prompts and responses.

The application of the signature approach to the CEQUEL data has revealed a previously
unobserved indicator of treatment response. We demonstrated that the behavioural patterns of subjects - their delay in responding to a prompt delivered by text message asking about their depressive symptoms -  are affected by treatment. The actual self-reported scale scores (QIDS-SR$_{16}$) are a subjective measure of the mood condition, while the delay patterns may be more objective correlate of mood. More broadly, the findings highlight the value of digital data capture which allows us to explore hitherto unknown or unmeasurable information. The collection of metadata is complemented by new mathematical approaches which can reveal new metrics and allow us to exploit low friction methods for data capture, which are of rapidly increasing use by health care systems to improve diagnosis and disease monitoring. 

The small sample of 29 subjects does not allow us to generalise our findings to a larger population, but rather motivates us to initiate a new experimental study with main focus on response traits with proper data collection design.

\section*{Ethics}
CEQUEL was registered with EUdraCT, number 2007-004513-33; and approved by REC 08/H0605/39; with a clinical trial authorisation 20584/0234/001-0001 and ISRCTN17054996. 

\section*{Data accessibility}
Data can be requested from JRG. 

\section*{Competing interests}

PJH, ABK, TJL, KEAS and JRG declare no competing interests.

\section*{Authors' contributions}
JRG and PJH designed the trial. JRG coordinated the study. AK and TL conducted the analysis. ABK, TJL and KEAS drafted the paper, which was reviewed by all authors.

\section*{Acknowledgements}

ABK and TJL wish to thank the Oxford-Man Institute for hospitality.

\section*{Funding}
The CEQUEL study was funded by the Medical Research Council. Some study drug was donated by GlaxoSmithKline.
ABK, TJL, KEAS, JRG and PJH are supported by a Wellcome Trust strategic award "CONBRIO", 102616/Z/13/Z.
Neither funder had any role in the study design; data collection, analysis, or interpretation of data; writing of the report; or in the decision to submit the paper for publication.
TL acknowledges the support of NCEO project NERC, ERC grant number 291244 and EPSRC grant number EP/H000100/1.

\bibliographystyle{JHEP}
\bibliography{biblio}

\providecommand{\href}[2]{#2}\begingroup\raggedright\begin{thebibliography}{10}

\bibitem{lyons1998differential}
T.~J. Lyons, {\it Differential equations driven by rough signals},  {\em
  Revista Matem{\'a}tica Iberoamericana} {\bf 14} (1998), no.~2 215--310.

\bibitem{kloeden2012numerical}
P.~E. Kloeden, E.~Platen, and H.~Schurz, {\em Numerical solution of SDE through
  computer experiments}.
\newblock Springer Science \& Business Media, 2012.

\bibitem{kusuoka2004approximation}
S.~Kusuoka, {\it Approximation of expectation of diffusion processes based on
  lie algebra and malliavin calculus},  in {\em Advances in mathematical
  economics}, pp.~69--83.
\newblock Springer, 2004.

\bibitem{lyons2004cubature}
T.~Lyons and N.~Victoir, {\it Cubature on wiener space},  in {\em Proceedings
  of the Royal Society of London A: Mathematical, Physical and Engineering
  Sciences}, vol.~460, pp.~169--198, The Royal Society, 2004.

\bibitem{chen1957integration}
K.-T. Chen, {\it Integration of paths, geometric invariants and a generalized
  baker-hausdorff formula},  {\em Annals of Mathematics} (1957) 163--178.

\bibitem{anderson2012bipolar}
I.~M. Anderson, P.~M. Haddad, and J.~Scott, {\it Bipolar disorder},  {\em Bmj}
  {\bf 345} (2012) e8508.

\bibitem{judd2003prospective}
L.~L. Judd, H.~S. Akiskal, P.~J. Schettler, W.~Coryell, J.~Endicott, J.~D.
  Maser, D.~A. Solomon, A.~C. Leon, and M.~B. Keller, {\it A prospective
  investigation of the natural history of the long-term weekly symptomatic
  status of bipolar ii disorder},  {\em Archives of General Psychiatry} {\bf
  60} (2003), no.~3 261--269.

\bibitem{judd2002long}
L.~L. Judd, H.~S. Akiskal, P.~J. Schettler, J.~Endicott, J.~Maser, D.~A.
  Solomon, A.~C. Leon, J.~A. Rice, and M.~B. Keller, {\it The long-term natural
  history of the weekly symptomatic status of bipolar i disorder},  {\em
  Archives of general psychiatry} {\bf 59} (2002), no.~6 530--537.

\bibitem{goodwin2003suicide}
F.~K. Goodwin, B.~Fireman, G.~E. Simon, E.~M. Hunkeler, J.~Lee, and D.~Revicki,
  {\it Suicide risk in bipolar disorder during treatment with lithium and
  divalproex},  {\em Jama} {\bf 290} (2003), no.~11 1467--1473.

\bibitem{geddes2016comparative}
J.~R. Geddes, A.~Gardiner, J.~Rendell, M.~Voysey, E.~Tunbridge, C.~Hinds, L.-M.
  Yu, J.~Hainsworth, M.-J. Attenburrow, J.~Simon, et~al., {\it Comparative
  evaluation of quetiapine plus lamotrigine combination versus quetiapine
  monotherapy (and folic acid versus placebo) in bipolar depression (cequel): a
  2$\times$ 2 factorial randomised trial},  {\em The Lancet Psychiatry} {\bf 3}
  (2016), no.~1 31--39.

\bibitem{rush200316}
A.~J. Rush, M.~H. Trivedi, H.~M. Ibrahim, T.~J. Carmody, B.~Arnow, D.~N. Klein,
  J.~C. Markowitz, P.~T. Ninan, S.~Kornstein, R.~Manber, et~al., {\it The
  16-item quick inventory of depressive symptomatology (qids), clinician rating
  (qids-c), and self-report (qids-sr): a psychometric evaluation in patients
  with chronic major depression},  {\em Biological psychiatry} {\bf 54} (2003),
  no.~5 573--583.

\bibitem{bennabi2013psychomotor}
D.~Bennabi, P.~Vandel, C.~Papaxanthis, T.~Pozzo, and E.~Haffen, {\it
  Psychomotor retardation in depression: a systematic review of diagnostic,
  pathophysiologic, and therapeutic implications},  {\em BioMed research
  international} {\bf 2013} (2013).

\bibitem{buyukdura2011psychomotor}
J.~S. Buyukdura, S.~M. McClintock, and P.~E. Croarkin, {\it Psychomotor
  retardation in depression: biological underpinnings, measurement, and
  treatment},  {\em Progress in Neuro-Psychopharmacology and Biological
  Psychiatry} {\bf 35} (2011), no.~2 395--409.

\bibitem{american2013diagnostic}
A.~P. Association et~al., {\em Diagnostic and statistical manual of mental
  disorders (DSM-5{\textregistered})}.
\newblock American Psychiatric Pub, 2013.

\bibitem{harrison2016innovative}
P.~J. Harrison, A.~Cipriani, C.~J. Harmer, A.~C. Nobre, K.~Saunders, G.~M.
  Goodwin, and J.~R. Geddes, {\it Innovative approaches to bipolar disorder and
  its treatment},  {\em Annals of the New York Academy of Sciences} {\bf 1366}
  (2016), no.~1 76--89.

\bibitem{levin2013learning}
D.~Levin, T.~Lyons, and H.~Ni, {\it Learning from the past, predicting the
  statistics for the future, learning an evolving system},  {\em arXiv preprint
  arXiv:1309.0260} (2013).

\bibitem{gyurko2013extracting}
L.~G. Gyurk{\'o}, T.~Lyons, M.~Kontkowski, and J.~Field, {\it Extracting
  information from the signature of a financial data stream},  {\em arXiv
  preprint arXiv:1307.7244} (2013).

\bibitem{lyons2011inversion}
T.~Lyons and W.~Xu, {\it Inversion of signature for paths of bounded
  variation},  {\em arXiv preprint arXiv:1112.0452} (2011).

\bibitem{lyons2007differential}
T.~J. Lyons, M.~Caruana, and T.~L{\'e}vy, {\em Differential equations driven by
  rough paths}.
\newblock Springer, 2007.

\bibitem{hambly2010uniqueness}
B.~Hambly and T.~Lyons, {\it Uniqueness for the signature of a path of bounded
  variation and the reduced path group},  {\em Annals of Mathematics} (2010)
  109--167.

\bibitem{yang2015character}
W.~Yang, L.~Jin, and M.~Liu, {\it Character-level chinese writer identification
  using path signature feature, dropstroke and deep cnn},  {\em arXiv preprint
  arXiv:1505.04922} (2015).

\bibitem{chevyrev2016primer}
I.~Chevyrev and A.~Kormilitzin, {\it A primer on the signature method in
  machine learning},  {\em arXiv preprint arXiv:1603.03788} (2016).

\bibitem{kiraly2016kernels}
F.~J. Kir{\'a}ly and H.~Oberhauser, {\it Kernels for sequentially ordered
  data},  {\em arXiv preprint arXiv:1601.08169} (2016).

\bibitem{yin2013icdar}
F.~Yin, Q.-F. Wang, X.-Y. Zhang, and C.-L. Liu, {\it Icdar 2013 chinese
  handwriting recognition competition},  in {\em Document Analysis and
  Recognition (ICDAR), 2013 12th International Conference on}, pp.~1464--1470,
  IEEE, 2013.

\bibitem{tibshirani1996regression}
R.~Tibshirani, {\it Regression shrinkage and selection via the lasso},  {\em
  Journal of the Royal Statistical Society. Series B (Methodological)} (1996)
  267--288.

\bibitem{zou2005regularization}
H.~Zou and T.~Hastie, {\it Regularization and variable selection via the
  elastic net},  {\em Journal of the Royal Statistical Society: Series B
  (Statistical Methodology)} {\bf 67} (2005), no.~2 301--320.

\bibitem{chawla2002smote}
N.~V. Chawla, K.~W. Bowyer, L.~O. Hall, and W.~P. Kegelmeyer, {\it Smote:
  synthetic minority over-sampling technique},  {\em Journal of artificial
  intelligence research} (2002) 321--357.

\bibitem{he2008adasyn}
H.~He, Y.~Bai, E.~A. Garcia, and S.~Li, {\it Adasyn: Adaptive synthetic
  sampling approach for imbalanced learning},  in {\em Neural Networks, 2008.
  IJCNN 2008.(IEEE World Congress on Computational Intelligence). IEEE
  International Joint Conference on}, pp.~1322--1328, IEEE, 2008.

\bibitem{kohavi1995study}
R.~Kohavi et~al., {\it A study of cross-validation and bootstrap for accuracy
  estimation and model selection},  in {\em Ijcai}, vol.~14, pp.~1137--1145,
  1995.

\bibitem{cawley2010over}
G.~C. Cawley and N.~L. Talbot, {\it On over-fitting in model selection and
  subsequent selection bias in performance evaluation},  {\em The Journal of
  Machine Learning Research} {\bf 11} (2010) 2079--2107.

\bibitem{gyurko2008rough}
L.~G. Gyurko and T.~Lyons, {\it Rough paths based numerical algorithms in
  computational finance}, .

\bibitem{WinNT}
S.~Buckley, L.~G. Gyurko, C.~Litterer, A.~Janssen, D.~Chafai, T.~Lyons, and
  A.~Papavasiliou, ``Computational rough paths.''

\end{thebibliography}\endgroup

\end{document}